\documentclass[11pt]{article}
\usepackage{moriond,epsfig}

\def\Journal#1#2#3#4{{#1} {\bf #2}, #3 (#4)}

\def\aa{\em A\&A}
\def\aas{\em A\&AS}
\def\apj{\em ApJ}
\def\mn{\em MNRAS}
\def\nat{\em Nature}

\newcommand{\crocho}{\left[\begin{array}{cc}}
\newcommand{\crochf}{\end{array}\right]}
\newcommand{\fb}{f_{\rm b}}
\newcommand{\fg}{f_{\rm gas}}
\newcommand{\h}{\:h_{50}}

\newcommand{\mb}{M_b}

\newcommand{\mprot}{m_{\rm p}}

\newcommand{\mst}{M_*}

\newcommand{\mtot}{M_{\rm tot}}
\newcommand{\omo}{\Omega_{\rm 0}}
\newcommand{\omb}{\Omega_{\rm b}}
\newcommand{\paro}{\left(\begin{array}{cc}\!\!}
\newcommand{\parf}{\!\!\end{array}\right)}

\newcommand{\rcx}{r_{\rm cX}}

\newcommand{\tx}{T_{\rm X}}

\def\be{\begin{equation}}
\def\ee{\end{equation}}
\def\bea{\begin{eqnarray}}
\def\eea{\end{eqnarray}}

\begin{document}
\vspace*{4cm}
\title{THE BARYON FRACTION DISTRIBUTION IN X-RAYS GALAXY CLUSTERS
}

\author{ R. SADAT \& A. BLANCHARD }

\address{Observatoire de midi-pyr\'en\'ees, 14 avenue Edouard Belin- 31400 Toulouse, France}
\maketitle\abstracts{
The baryon fraction in galaxy clusters is one of the most direct way to constrain $\omo$. The baryonic fraction is estimated to be in the range $\fb = (0.15 - 0.20)\h^{-3/2}$ which is several times higher than expected from the observed light element abundances in $\omo=1$ universe, leading to the conclusion that we live in low density universe.
In this work I will first present the various steps which lead to these results and then foccus on the distribution of the baryon fraction inside clusters. I will show that while the baryon fraction in clusters follows a scaling law, its theoretical expected shape does not agree with what is inferred from observations. I will show that when various factors entering into the determination of clusters gas fraction are taken into account the observed baryon fraction profile matches better the predictions, provided that the average baryon fraction at the virial radius is lower than previously found $\fb \sim 10 \%$.}

\section{Introduction}

The baryon fraction in clusters, $\fb$, is assumed to be equal to the cosmic value. Combined with the universal baryon density $\omb$ as predicted from light element abundances through the theory of big-bang nucleosynthesis it yields to the upper limit of the cosmic mass density: 

\begin{equation}
\omo= \Gamma(r_{500})\frac{\omb}{\fb(r_{500})}
\end{equation}
The numerical factor $\Gamma$ ($\sim 0.92$) is a correction factor for possible differences between the baryon content in clusters and the universal one arising during cluster formation. There are still uncertainties on the current BBN results on the primordial D/H abundance with $0.024 \le \omb\h^{2} \le 0.08$. As the formulae shows, this test to $\omo$ is in principle the most direct one as it depends on very few reasonable assumptions: the cluster formation by collapse from a well mixed medium, and no segregation between the gas and the dark matter as shown by numerical simulations.\\
Rosat observations of Coma cluster have led to the following numbers: Baryonic content : $\mb \sim 3.4\h^{-5/2}10^{14}M_{\odot}$; Mass in stars : $\mst \sim 2.2\h^{-1}10^{13}M_{\odot}$; Total mass : $\mtot \sim 2.2 \h^{-1}10^{15}M_{\odot}$. These numbers yields to a gas fraction of $\fg \sim 0.15 \h^{-3/2}$ at Abell radius. When combined to current estimates of $\omb \h^{2}$ this leads to low value of $\omo$ (White et al. 1993). Several analyses of large samples of clusters have confirmed such a high baryon fraction with values ranging from $15-20\%$. Hower, there is large dispersion between the published values and it has even been argued that the baryon fraction increases with cluster temperature (mass) and with radius inside a cluster. This fact has often been explained as due to non-gravitational processes such as heating effects occuring during galaxy cluster formation. In this talk I will address the issue of the baryon fraction in clusters by examining critically its distribution within the cluster and I will show how some systematic effects can bias the baryon fraction determination. 
\section{The standard method of gas fraction determination}
The main assumption of this method is that the galaxy baryon fraction reflects the universal one.
The main component of the baryonic content is mostly in the form of hot X-ray gas. The standard way to derive the gas fraction proceeds from the following steps:\\
\noindent
$\bullet$ Assume a $\beta-model$ to fit the observed surface brightness $S_{o}(1+(\theta/\theta_{c})^2)^{-3\beta+1/2}$ where $\theta$ is the projected angular distance to the center. The gas density profile (thus the gas mass profile) is then derived using the two best-fit parameters $\beta$ and $\theta_{c}$. The X-ray gas emission is rarely traced out to very large radii, so that gas fractions are measured only up to an X-ray limiting radius for which the signal-to-noise is good enough. Most of  the time, it was necessary  to extrapolate to the virial radius $r_{200}$ or to some other outer radius like $r_{500}$. It is then essential to realize that gas mass is estimated at a radius at which the actual emission is poorly constrained, with a value of the order of the X-ray background or less. Moreover, the emissivity being dominated by the central region, a parametric fit will be rather insensitive to the outer part of the gas profile.\\
$\bullet$ Use the hydrostatic equation to derive the total mass inside the same radius: $\mtot(r)=\frac{3k}{G \mu \mprot} \beta \tx r \paro 1 +
      \paro \frac{r}{\rcx} \parf ^{-2} \parf ^{-1}$.
The total mass thus depends linearly on both $\beta$ and $\tx$.
Hence, if the slope of the gas density is poorly determined, it will have a drastic influence on the
derived mass. 
\section{The baryon fraction profile: Observations versus simulations}
\begin{figure}[htbp]
\resizebox{6.5cm}{!}{\includegraphics{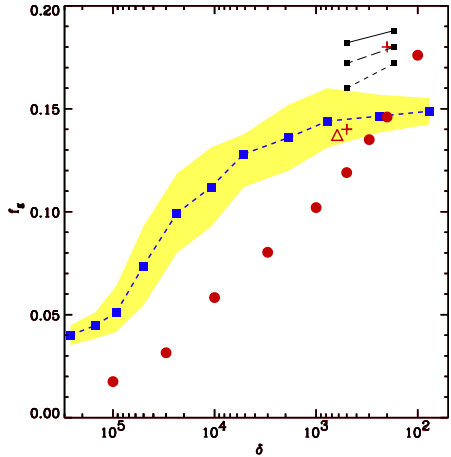}}
\hfill
\parbox[b]{90.5mm} 
{\caption{ \small The distribution of the gas fraction $\fg$ versus $\delta$. The observations are from RSB00 
 {\it filled circles}, Ettori \& Fabian 1999 {\it empty triangle} and Arnaud \& Evrard 1999 {\it crosses}. Statistical uncertainties on these quantities are smaller than the symbols size. The {\it squares} connected by a 
dashed line correspond to the theoretical gas fraction for a global value of $16\%$ (from the numerical simulations of the Santa Barbara group). {\it Small squares} connected by a line correspond to the gas fraction calculated from 
numerical simulations including winds by Metzler \& Evrard (1997): for a 6 keV cluster ({\it long dashes}), a 3 keV cluster ({\it small dashes}) and without including winds ({\it continuous line}) with a global value of 20\%. }
 \label{fig:fbobs}}
\end{figure}

In the case where only gravity is acting during cluster formation, the baryon (gas) fraction profile is expected to follow a scaling law, depending only on the contrast density $\delta$ ($\delta \equiv \rho(<r)/\rho_c$ where 
$\rho_c \equiv 3H^{2}_{o}/8\pi G$). Hydro-dynamical simulations of an X-ray cluster from different groups (the Santa Barbara group) have shown that the gas fraction normalized to the global value, is a function of radius it first increases in the inner part and then tends to flatten in the outer part to reach the (cosmic) value see Fig. \ref{fig:fbobs} . The distribution of the gas fraction within clusters has been widely examined by Roussel et al. (2000) (hereafter RSB00) for a large sample of groups and galaxy clusters. The resulting shape is shown in (Fig. \ref{fig:fbobs}). To ensure consistency I will use the scaling law derived gas fractions by RSB00, in which masses are derived from the NFW universal dark matter profile normalized to the $M_{\delta} - \tx$ relation from numerical simulations with two calibrations: $\tx = 4.75 (M_{200}/10^{15}M_{\odot})^{2/3} \textnormal{keV}$ (EMN96) and $\tx = 3.81 (M_{178}/10^{15}M_{\odot})^{2/3}$ \textnormal{keV} (Bryan and Norman 1998, BN98), $M_{\delta}$ being the mass enclosed in a region with a density equal to 
 $(1+\delta)$ times the critical density. These two values of the normalization can be considered to be the extreme values among existing numerical simulations. BN98 normalization leading to virial masses nearly 40\% higher.
 For comparison gas fractions at $r_{500}$ and $r_{200}$ from the literature are also plotted in (Fig. \ref{fig:fbobs}). 
The comparison of both profiles is very surprising as it shows that:
the gas fraction profile derived from observations is in strong disagreement with the numerical simulations results (for a global value of $16\%$), its shape continuously increases and does not exhibit the flattening in the outer parts as seen in numerical simulations.
Clearly, this discrepancy calls for caution when one is using the gas fraction to set upper limit on the mean density of the universe.\\
One possibility is that processes acting during the cluster formation are not well understood and probably non-gravitational processes could have played an important role. But this is probably not the case, because in that case large departure from scaling laws is expected which is not observed in the data (RSB00).
\section{Uncertainties in gas fraction estimation}
\subsection{The extrapolation effect}
In a recent study of a large sample of X-ray ROSAT images in which the data trace the emission up to very large radii (approximately up to the virial radius), it has been shown that the $\beta$-model does not provide an accurate description of the surface brightness over the whole range of radii. They have found that the outer slopes (0.3$r_{200}$ to $r_{200}$) are actually steeper than the slope found when the $\beta$-model is applied to the whole cluster which actually leads to lower gas masses (Vikhlinin et al. 1999, VFJ99). 
\subsection{The clumping effect}
In most of the studies on gas and total mass estimations, the ICM gas is assumed to be uniform, which is not true, as actual clusters do exhibit a certain level of density fluctuations at small and large scale  probably due to accretion and mergers events. This clumping is assumed to be more pronouced in the outer parts of the clusters where the relaxation is not completely achieved. In the presence of clumping the gas masses are overestimated by a factor $\sqrt{C}=1.16$at $\delta = 500$ ($C = <\rho_g^2>/<\rho_g>^2$) (Mathiesen et al. 1999, MEM99).\section{Constraining $\omo$ using the corrected gas fraction}
\begin{figure}[htbp]
\resizebox{6.9cm}{!}{\includegraphics{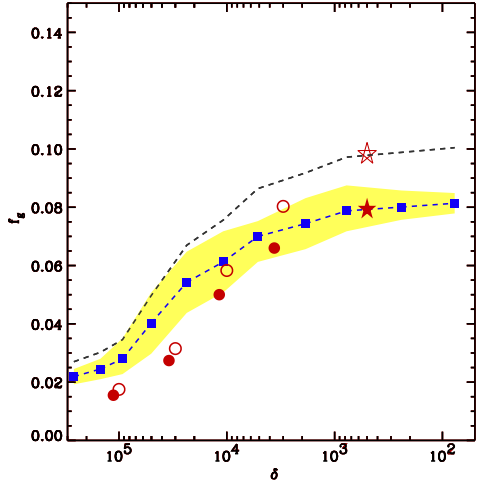}}
\hfill
\parbox[b]{80.5mm} 
{\caption{\small The distribution of the gas fraction $f_g$ versus $\delta$: {\it circles} correspond to the observed $f_g$ from RSB00 but restricted to the region where the gas is detected (no extrapolation), using EMN96 ({\it empty circles}) and BN98 ({\it filled circles}) calibration. {\it  Empty} ({\it filled}) {\it star} corresponds to the estimated $f_g$ using VFJ99 sample and corrected for the clumping with EMN96 (BN98) calibration. {\it Filled squares} linked by a dashed line correspond to the gas fraction predicted by numerical simulations for a global fraction of 8.75\%. The upper dashed line corresponds to the case where the global fraction is 10.8\% }
 \label{fig:fb_res}}
\end{figure}

I will now estimate the new gas fractions taking into account the above effects. The gas masses are corrected from the extrapolation problem by using the VFJ99 sample (we use their results to compute the gas mass in the virial region) and for the clumping effect using the MEM99 factor. Virial masses are computed from numerical simulations as described in section 3. The result is plotted in (Fig. \ref{fig:fb_res}). From this figure, we can see that the gas fraction distribution, in particular the asymptotic behaviour, is now consistent with numerical simulations predictions, (although a slight difference persists in the inner parts at $\delta <10^{4}$ which can well be due to non-negligible energy injection in these regions). By matching the data point at $\delta=500$, we find $\fg=0.0875 \pm 0.0075$ ($0.108\pm0.0092$) at $68\%$ confidence level with BN98 (EMN96) normalization, the uncertainty being due to numerical simulations (Fig. \ref{fig:fb_res}). Adding the stellar mean contribution of $1\%$, our final value of the baryon fraction at the virial radius is:  $\fb \sim 0.10 \pm 0.01$ (with BN98 calibration) and $\fb \sim 0.12 \pm 0.01$ (with EMN96 calibration).
Plugging this new value of the baryon fraction in equation (1) and using the low D/H value as recently reported by O'Meara et al. (2000) corresponding to $\omb \sim 0.08$ (this value is consistent with recent results on CMB fluctuations measurements from Boomerang mission 2001), we find the following value for $\omo$:
$\omo = 0.8 \pm 0.1$ 
using the baryon fraction estimated with BN98 normalization. This value is consistent with the result from the evolution of X-ray clusters abundance (Blanchard et al. 2000). If we use the baryon fraction obtained with EMN96 calibration, we find $\omo \sim 0.65$.

\section{conclusion}
The baryon fraction in X-ray clusters is one of the most direct way to constrain $\omo$. Previous estimations of the baryon fraction yield to a high value $\fb \sim 15\% \h^{-3/2}$, favouring low density universe $\omo \sim 0.3$. In my talk I have shown that the radial profil of the gas fraction as derived from observations differs strongly from what is found in numerical simulations. I argue this is not due to non-gravitational processes which may occur during the cluster formation but is rather due to various systematics such as the extrapolation of a $\beta$-model and the clumpiness of the gas that have not been taken into account in previous estimations. Correcting the gas masses from these biases and using total masses calibrated from numerical simulations yield to a better agreement between observations and numerical simulations for a global value of the order of $\sim 10 (12)\%\h^{-3/2}$ depending on which normalization we use to compute the total mass. This revised value is lower than previous ones and leads to $\omo \sim 0.8 (0.65)\h^{-1/2}$. \\

\end{document}